\documentclass[12pt,preprint]{aastex}

\begin{document}
\title{Influence of Planets on Parent Stars: Angular Momentum}
\author{Noam Soker}
\affil{Department of physics, Technion$-$Israel Institute of Technology,
Haifa 32000, Israel, and Department of Physics,
Oranim; soker@physics.technion.ac.il}

\begin{abstract}
I review some possible processes by which planets and brown dwarfs
can influence the evolution of their parent evolved stars.
As sun-like stars evolve on the red giant branch (RGB)
and then on the asymptotic giant branch (AGB), they will
interact with their close planets (if exist).
The interaction starts with tidal interaction: this will lead the
planets to deposit most of their angular momentum to the
envelope of the giant, and then spiral-in to the envelope.
(Too many papers dealing with close planets [less than about
3-6 AU] around evolved stars neglect tidal interaction,
hence their results are questionable.)
They may spin-up their parent stars by up to several
orders of magnitude.
The interaction of substellar objects with evolved star
may enhance the mass loss rate, mainly in the equatorial plane.
Possible outcomes are: ($i$) Planetary systems interacting with
their parent AGB star may lead to the formation
of moderate elliptical planetary nebulae.
($ii$) RGB stars which lose more mass turn to bluer
horizontal branch (HB) stars. Therefore, planets may explain the
formation of blue HB stars. This may explain the
presence of many blue HB stars in many globular clusters
(the planets be the {\it second parameter}), and some hot
HB stars in the galaxy (sdB stars).
(The 8.3 days use of the Hubble Space Telescope in search of
planets in a globular clusters with no blue HB stars was a
wrong move.)
($iii$) Most known stars with planets will not form planetary
nebulae, because they will lose most of their envelope
already on the RGB.

\end{abstract}

\section{Introduction}

The evolution of planets inside the envelopes of evolved stars
was studied long before the detection of extrasolar planets
(e.g., Eggleton 1978; Livio 1982; Livio \& Soker 1984;
Harpaz \& Soker 1994). 
This is also true for the study of the  influence of planets and 
brown dwarfs on their parent stars.  
One of the main aspects of motivation was to explain the formation of
some moderate axisymmetrical, i.e., elliptical, planetary nebulae 
(PNs; bipolar and extreme elliptical PNs require
stellar binary companions; Soker 1997; 1998b; 2002b). 
The immediate progenitors of PNs, which are asymptotic giant
branch (AGB) stars, are expected to rotate extremely slowly.
Even an earth-like planet can spin-up its parent
star if it enters the envelope as the star is about to leave the AGB;
Jupiter-like planets may spin-up their parent upper-AGB stars by
several orders of magnitude (section 5).
Planets were also suggested to enhance mass loss rate from
stars on the red giant branch (RGB), such that when the star
turns to the horizontal branch (HB) on the HR diagram,
it will have low-mass envelope and be a blue HB star
(section 4).
As the planet is destructed inside the giant envelope, it may
change its composition, e.g., deposits lithium.
Other effects due to planets were also suggested.
For example, planets outside the envelope may accrete mass and have 
bursts of emission (Struck, Cohanim, \& Willson 2002, 2004), and may 
influence SiO maser emission (Struck-Marcell 1988; Struck et al.\ 2004),
as well as H$_2$O masers (Rudnitskij 2002). 
I think it will be very difficult to find planets accreting from
AGB stars: in order to accrete at a high rate, or to strongly interact
with the AGB wind, a low mass companion must be close
to the AGB surface;
but if it is too close, tidal interaction will cause it to spiral in  
during a short period of time (see section 3 below).
So only a small fraction of systems will
go through this stage, and even then the luminosity of the accreting
planet will be very low. 
Planets spiraling in should accrete at a high rate when
close to the envelope, and as was found by Struck et al.\ (2004),
may form an accretion disk
(I think Struck et al.\ [2004] overestimate the accretion rate of
angular momentum because they use a 2-dimensional rather than 
3-dimensional numerical code.) 
The plant then may blow two jets. 
The idea that planets may blow jets was raised in the context of
young planetary systems, where the planets accrete from the 
proto planetary system disk (Quillen \& Trilling 1998). 
In any case, my view is that there is no need to invoke direct
influence of planets to explain SiO maser variabilities, 
or replace pulsation with planets to explain oscillations of AGB
stars as suggested by Berlioz-Arthaud (2003); 
the pulsation and strong convection of AGB stars, with possible
weak magnetic activity, can account for these variabilities
in AGB stars.
(I think the results of some of these papers, which deal
with planets around AGB stars are also questionable 
because they ignore the importance of tidal forces
when the planets are close to the AGB envelopes.)
I think, though, that planets may have an indirect influence on 
them (see below). 

The main point I would like to make is that planets may have
pronounced effects on their parent stars, and by studying
them we may solve some puzzles in stellar evolution.
The first issue to be considered in this respect is the definition
of a single star. From the `official' point of view, a star
and a planet are a single-star system.
However, for someone simulating stellar evolution, this may not
be the case.
If the evolutionary code, or other means of calculating the evolution,
do not include relevant effects of planets around a specific object,
e.g., spinning up the envelope and enhancing mass loss rate,
causing mixing at the core-envelope interface,
depositing fresh hydrogen-rich material to the nuclear burning shell,
etc., wrong results will be obtained. For this specific case, the
star is not really a single star, although it has no stellar-companions.
In the past I have suggested that {\it a non-single star will be one
for which one of its relevant properties, e.g., angular momentum,
hydrogen abundance in the core, etc., is determined by a
gravitationally bound object.}

\section{Nonlinear Effects of Planets}

Enhanced equatorial mass loss rate due to centrifugal forces is
not important for envelope spun-up by planets.
In addition, the gravitational energy of mass accreted onto the
planets are in general not sufficient in order to change the
structure of the descendant PN.
Therefore, there is a need for non-linear effects.
By nonlinear effects I refer to effects that are very sensitive to
the tiny effect of planets.

\subsection{Excitation of Waves in Common Envelopes}

One such nonlinear effect is the excitation of p-waves
(Soker 1992a, 1993) and g-waves (Soker 1992b)
during a common envelope phase.
While inside the convective envelope of an AGB or RGB star,
a companion will excite p-waves which propagate outward with
increasing amplitude, mainly in the equatorial plane.
For a planet of mass $\sim 10 M_J$, where $M_J$ is the Jupiter mass,
the surface pressure amplitude $P^\prime$ is (Soker 1993, eq. 4.1)
$P^\prime/P \sim 0.4$ for a secondary at $a_2 \simeq 0.1 R_\ast$,
where $P$ is the average surface pressure, $a_2$ is the location
of the companion from the center, and $R_\ast$ is the AGB
stellar radius. The perturbation increases linearly with the companion
mass (for low mass companions), and increases somewhat
as $a_2$ decreases (depending on convective viscosity). 
The amplitude is much larger in the equatorial plane than in the
polar directions.
Such excited non-radial oscillation can enhance
mass loss rate in the equatorial plane.

\subsection{Destruction of Planets in the Envelope}

A study of the fate of planets in the envelope of AGB stars
was conducted by Livio \& Soker (1984). They assumed that the
planet accretes from the envelope at the Bondi-Hoyle accretion
rate.
However, it is possible that the planet swells as a
result of this accretion and does not accrete much,
like low-mass main sequence stars do (Hjellming \& Taam 1991).
Planets may also be evaporated, in particular when they
reach the place in the envelope where the envelope's temperature
exceeds the planet's virial temperature.
For RGB and AGB stars, the orbital separation of a planet
from the core where fast evaporation starts is (Soker 1998a)
\begin{equation}
a_2({\rm evaporation}) \simeq 10
\left( \frac{M_p}{M_J} \right)^{-1} R_\odot,
\end{equation}
where $M_p$ is the planet's mass.
The cool and dense evaporated material is still of low entropy,
and fraction of it may spiral-in to the core.
More massive planets than Jupiter will survive farther in, until
they reach a radius where Roche lobe overflow (RLOF) occurs.
For a planet of radius $R_p=0.1 \eta R_\odot$, RLOF occurs when
the orbital separation from the core is (see Soker 1998a)
\begin{equation}
a_2({\rm RLOF}) \simeq 1.7 \eta
\left( \frac{M_p}{M_J} \right)^{-1/3} R_\odot.
\end{equation}

The supplement of the destructed planet (or brown dwarf) 
material to the core and around it may have several effects.
First, the low entropy material can absorb heat, and
may reduce for a short period of time the stellar luminosity
(Harpaz \& Soker 1994). If the material reaches the core,
or close to it, the release of gravitational energy and nuclear burning
of the fresh hydrogen-rich material may lead to stellar expansion
and enhanced mass loss rate (Siess \& Livio 1999a,b).
Recently, Retter \& Marom (2003) proposed that three planets 
which deposited gravitational, and then nuclear, energy into
their parent RGB star, along the calculations of Siess \& Livio (1999b),
with about a month delay from one planet to the next, can account for the
erupting of V838 Mon. 
Second, the high specific angular momentum of the planet's
(or brown dwarf) material may lead to the formation of an accretion 
disk around the core; such disk can launch two jets (Soker 1996b).

\subsection{Spinning-up RGB and AGB Envelopes and Magnetic Activity}

The spinning-up of RGB and AGB envelopes was discussed in 
several papers (see Soker 2001b). 
Basically, RGB and AGB stars slow down rapidly as they lose mass
(see figures 1 and 2 in Soker 2001b). Planets and brown dwarfs can 
then tidally interact with the expanding star (see next section),
enter the envelope, and deposit their orbital angular momentum
to spin-up the envelop by a factor up to $\sim 10^4$, depending on
the mass of the planet. 
However, the envelope will still spin much below the Keplerian
speed on the equator (the break-up speed). To influence the
mass loss process, a non-linear effect must be incorporated. 
Such an effect appears in the cool magnetic spots model (Soker 1998c;
Soker \& Clayton 1999),  where it is assumed that a weak magnetic 
field forms cool stellar spots, which facilitate the formation of 
dust closer to the stellar surface, hence increasing the mass loss
rate.
If magnetic spots, due to the dynamo activity, are formed mainly near
the equator, an enhanced equatorial mass loss is obtained.
The weak magnetic field is assumed to be formed by the strong
convection in AGB and RGB stars, together with a very slow rotation,
which mainly serves for defining the symmetry axis of the magnetic activity. 

\section{Tidal Interaction}

RGB and AGB stars tidally interact with their companion before
RLOF or common envelope occur. Most of the orbital angular momentum
of the companion is deposited to the envelope before the
onset of the common envelope phase. Therefore, the star can
lose substantial fraction of its mass before the common
envelope phase starts (Soker 2002c; Soker \& Harpaz 2003). 
During the evolution along the RGB or AGB, the star expands,
and tidal interaction strength increases steeply with the
giant radius. 
It is mandatory to take into account tidal interaction,
with substellar or stellar companions, in studying the interaction
of giants with their close companions. 
The tidal interaction, for example, is crucial in determining the
fate of the earth as the Sun becomes an AGB (Rybicki \& Denis 2001;
see also Rasio et al.\ 1996).
Since I find that too many papers dealing with planets
around AGB and RGB stars ignore tidal interaction, hence
overestimating the importance of the effects they study,
I devote a section to this subject.

In Soker (1996a) I found the maximum orbital separation at which 
tidal interaction is significant. 
For planets and brown dwarfs, because they cannot bring the envelope 
to corotate, this radius is the radius below which they will spiral
into the envelope of the giant. This maximum radius is given by    
\begin{eqnarray}
a_{\rm max} = 3.9 R  
\left( {{\tau_{\rm ev}}\over{6\times10^5 ~{\rm yr}}} \right)^{1/8} 
\left( {{L} \over {2000 L_\odot}} \right)^ {1/24}
\left( {{R} \over {200 R_\odot}} \right)^ {-1/12} \nonumber \\
\times 
\left( {{M_{\rm {env}}} \over {0.5M_1}} \right)^ {1/8} 
\left( {{M_{\rm {env}}} \over {0.5M_\odot}} \right)^ {-1/24} 
\left( {{M_2} \over {0.01M_1}} \right)^ {1/8}, 
\end{eqnarray}
where $L$, $R$, and $M_1$ are the luminosity, radius, and
mass of the giant (RGB or AGB star), respectively, 
$M_{\rm {env}}$ is the giant's envelope mass, and $\tau_{\rm ev}$ 
is the evolution time on the upper AGB or RGB. 

\section{Influencing the Horizontal Branch Morphology}

 After RGB stars ignite helium in their core they move to the
horizontal branch (HB) on the HR diagram. 
Stars which lose more mass on the RGB become bluer (hotter) HB stars. 
The HB morphologies, i.e., the distribution of stars on the HB of 
a stellar system, differs substantially from one globular cluster 
to another.   It has long been known that metallicity is the main
factor which determines the location of HB stars on the HR diagram.
Metallicity is the  {\it first parameter}. 
For more than 30 years, though, it has been clear that another factor is
required to explain the variation in HB morphologies among 
globular clusters with similar metallicity 
(see reviews by  Fusi Pecci \& Bellazzini 1997; de Boer 1999).
This factor is termed the {\it second parameter} of the HB.
It seems that stellar companions alone cannot be the second parameter
(e.g., Rich et al.\ 1997), nor any other single factor which has
been examined (e.g., Ferraro et al.\ 1997 and references therein). 
I think that the presence of low mass stars and of planets
(or brown dwarfs) is the main second parameter factor (but probably not
the only one), with planets occurring more frequently (Soker 1998a).

In recent years it has become a common view that the second parameter
determines the HB morphology by regulating the mass loss on the
RGB  (e.g., Dorman, Rood, \& O'Connell 1993; D'Cruz et al.\ 1996, 2000;
Whitney et al.\ 1998; Catelan 2000). 
According to this view, the extreme HB (EHB) stars, for example, lose 
almost all their envelope  while still on the RGB
(Dorman et al.\ 1993; D'Cruz et al.\ 1996); mass loss on the
HB itself can't account for EHB stars (Vink 2003).
It is thought by many people that rotation has a connection with the 
second parameter through its role in determining the mass loss
on the RGB, directly or indirectly.
I agree with this assertion, and further claim that the source of
angular momentum in many cases is the interaction with a planet
(Soker \& Harpaz 2000; Livio \& Soker 2002).
Sweigart \& Catelan (1998 Moehler, Sweigart, \& Catelan 1999)
claim that mass loss on the RGB by itself cannot be
the second parameter, and it should be supplied by another process, 
e.g., rotation, or helium mixing, which requires rotation as well. 
They term the addition of such a process a ``noncanonical scenario''.
Behr et al.\ (2000b) find the second parameter problem to be
connected with rotation, and note that single star evolution cannot
explain the observed rotation of HB stars, even when fast core
rotations are considered.
The rich variety of HB morphologies (e.g.,  Catelan et al.\ 1998)
suggests that there is a rich spectrum in the factor(s) behind the
second parameter.

After presenting the idea that planets are the main factor in the
second parameter (Soker 1998a), I farther explored this idea in
three papers.
In Soker \& Harpaz (2000) we analyzed the angular momentum evolution
from the RGB to the HB and along the HB.
Using rotation velocities for stars in the globular cluster M13
(Behr et al.\ 2000b; similar distribution of rotation
is in the globular cluster M15; Behr, Cohen, \&  McCarthy 2000a),
we found that the required angular momentum for the fast rotators
is up to $1-3$ orders of magnitude (depending on some assumptions)
larger than that of the sun.
Planets of masses up to five times Jupiter's mass and up to an initial
orbital separation of $\sim 2$~AU are sufficient to spin-up the RGB
progenitors of most of these fast rotators. Other stars have been
spun-up by brown dwarfs or low-mass main sequence stars. 
Our results show that the fast rotating HB stars have been
probably spun-up by planets, brown dwarfs, or low-mass main
sequence stars, while they evolved on the RGB.

Support of the planet-second parameter idea comes from sdB
binary systems.
The field sdB stars are post-RGB stars, which have lost most of
their envelope, and are parallel to EHB (very hot) stars
in globular clusters (Stark \& Wade 2003; Vink 2003).
The class of objects named EC14026, which have sdB stars
and low mass main sequence companions, was discussed by, e.g.,
Kilkenny et al.\ (1997), Koen et al.\ (1997),
and Koen et al.\ (1998), and their relation to EHB
stars by (Bono \& Cassisi 1999).
PG 1336-018, for example, has a secondary of mass $\sim 0.15 M_\odot$
with an orbital period of $0.1$ days (Kilkenny et al.\ 1998).
Maxted et al.\ (2000) find the orbital periods and minimum
companion masses of two sdB stars: $0.63 M_\odot$ and $8.33$ days for
PG 0940+068, and $0.09M_\odot$ and $0.599$ days for PG 1247+554. 
For others, the companion, if it exists, is limited to a spectral type
M0 or later (e.g., PG 1605+072, Koen et al.\ 1998;
PG 1047+003, O'Donoghue et al.\ 1998).
For these systems, I suggest that the companion may be a brown dwarf
or a massive planet as well.
Allard et al.\ (1994) estimate that $\sim 60 \%$
of hot B subdwarfs have binary stellar companions.
Here again, the stars with no stellar companions may have a
substellar companion.
Green et al.\ (1998) argued that their ``investigations in open
clusters and the field strongly suggest that most
metal-rich BHB [blue HB] stars are in binaries''.
In many cases in the proposed scenario for the formation
of blue-HB stars the substellar companion will not survive
the evolution (Soker 1998a). Therefore, although most of the envelope
was lost as a result of the interaction with the substellar
companion, it does not exist anymore.

Soker \& Hadar (2001) study the correlations between the
the HB morphology and some other properties
of globular clusters.
Strengthening previous results, we find that a general correlation
exists only between HB morphology and metallicity.
Correlations with other properties, e.g., central density and
core radius, exist only for globular clusters within a narrow
metallicity range.
We conjecture that the lack of correlations with {\it present}
properties of globular clusters (besides metallicity), is because
the variation of the HB morphologies between
globular clusters having similar metallicities is caused by a
process, or processes, whose effect was determined
at the {\it formation time} of globular clusters.
This process (or processes) is the second parameter.
The `planet second parameter' model fits this conjecture.
This is because the processes which determine the presence
of planets and their properties occur during the formation
epoch of the star and its circumstellar disk.

Even if planets are not the second parameter, still, close
Jupiter-like planets must increase the mass loss from
RGB stars (Livio \& Soker 2002).
Therefore, if a large fraction of sun-like stars in a group,
like in a globular cluster, possesses close planets, many
of the HB stars in this group will be blue (hot).
Gilliland et al.\ (2000) used the HST for 8.3 days and
found no planets around main sequence stars in the globular
cluster 47 Tucanae (NGC 104).
This globular cluster contains only a few blue HB stars
(Rich et al.\ 1997; Moehler, Landsman \& Dorman 2000), 
and therefore I do not expect the stars in this globular
cluster to have massive and close planets around them.
Therefore, 47 Tuc was a bad choice for planets search.
(In my talk at the meeting I clearly made that point.
Despite that, in a paper to these proceedings [posted on astro-ph]
a new [ground] search for planets in 47 Tuc is presented,
without referring to my point.) 
In the globular cluster M4, on the other hand, close to
half of the HB stars are blue (Harris 1996), and indeed,
the presence of a planet in this globular cluster
was confirmed recently (Sigurdsson et al.\ 2003).
Ferdman et al.\ (2003) found no candidates for planet transits
in their HST study of M4. 
I suspect their study was not sensitive enough to eliminate
the possibility of planets in M4.
I encourage more HST observations of M4 and other globular clusters
having large population of blue-HB stars.

\section{Influencing the Mass Loss Geometry}

As mentioned in section 1, in several papers I examined
the possible role of planets in influencing the mass loss
geometry from AGB stars, with the goal of explaining
moderate elliptical PNs. 
The main process (section 2.3) is the dynamo amplification
of magnetic fields in slowly rotating AGB stars. This leads to
the formation of cool magnetic spots, which enhance 
dust formation and mass loss, mainly from the equatorial plane. 
The most recent papers, where more references can be found,
are Soker (2001b) and Livio \& Soker (2002).

In Soker (2001b) I examine the implications of the recently 
found extrasolar planets on the planet-induced axisymmetric 
mass loss model.
Since about half of all planetary nebulae are elliptical, i.e.,
have low equatorial to polar density contrast,
it was predicted that about $50 \%$ of all solar-like stars
have Jupiter-like planets around them, i.e., a mass about equal
to that of Jupiter, $M_J$, or more massive.
In  light of the new findings that only $5 \%$ of sun-like
stars have such planets, and the mechanism of dust formation 
near cool magnetic spots, I revise this prediction.
In Soker (2001b) I predict that indeed $\sim 50 \%$ of PNs 
progenitors do have close planets around them, but the 
planets can have much lower masses, as low as $\sim 0.01 M_J$, 
in order to efficiently spin-up the envelopes of AGB stars.
To support this claim, I follow the angular momentum evolution of
single stars with main-sequence mass in the range of 
$1.3-2.4 M_\odot$, as they evolve to the post-AGB phase.
I find that single stars  rotate much too slowly to possess any
significant non-spherical mass loss as they reach the upper AGB.
It seems, therefore, that planets, in some cases even Earth-like 
planets, are sufficient to spin-up the envelope of these 
AGB stars for them to form elliptical PNs.
The prediction that on average several such planets orbit each star,
as in the solar system, still holds. 
In Soker (2001a) I show that wide stellar companions to
AGB stars may also accrete mass, form an accretion disk,
and blow jets, hence forming elliptical PNs. 
This reduces the fraction of PN progenitors which are 
needed to have planetary systems from $\sim 50 \%$ to 
$\sim 35 \%$, or even less. 

Another finding from exoplanets is that metal rich stars 
are more likely to harbor planetary systems. 
This implies, in the context of planet-shaping of PNs,
that spherical PNs will tend to originate from
low metallicity stars.  
Indeed, when carefully defining spherical PNs, this is the case.
In Soker (2002a) I examine the mass loss history and 
distribution of spherical PNs in the galaxy.
I argue there that spherical PNs form a special group among all PNs.
The smooth surface brightness of most spherical PNs suggests
that their progenitors did not go through a final intensive
wind (FIW, also termed superwind) phase.
While $\sim 70 \%$ of the PNs of all other PNs groups are closer
to the galactic center than the sun is, only $\sim 30 \%$ of
spherical PNs are; $\sim 70 \%$ of them are farther away from the
galactic center.
These, plus the well known high scale height above
the galactic plane of spherical PNs, suggest that the progenitors 
of spherical PNs are low mass stars having low metallicity.

I also examine the possibility of detecting signatures 
of surviving Uranus, Neptune-like planets inside PNs (Soker 1999). 
Giant planets that are not too close to the stars (orbital separation 
larger than $\sim 5$~AU) are likely to survive the entire evolution of 
the star.  As the star turns into a PN, it has a fast wind and strong 
ionizing radiation. The interaction of the radiation and wind with a 
planet may lead to the formation of a compact condensation or tail 
inside the PN, which emits strongly in H$\alpha$, but not in [OIII]. 
The position of the condensation (or tail) will change over a 
time-scale of $\sim 10$~yr. Such condensations might be 
detected with currently existing telescopes. 
This idea was then repeated for planets around white dwarfs
(Chu et al. 2001). 

Finally, I note that most of the known stars with extrasolar
planets will {\it not} form PNs at all.
Instead, because they have relatively low mass (most have
$M<1.3 M_\odot$), I expect them to lose most of their envelope
on the RGB, becoming blue-HB stars, and then fading as WD without
an observable nebula.

\acknowledgments
This research was partly supported by the
Israel Science Foundation.

\end{document}